\documentclass[9pt,twocolumn,twoside]{osajnl}
\journal{ol} % Choose journal (ao,jocn,josaa,josab,ol,optica,pr)

\setboolean{shortarticle}{true}
\title{Electrically Poled Vapor-Deposited Organic Glasses for Integrated Electro-Optics}

\author[1]{Lauren Dallachiesa,}
\author[1,*]{Ivan Biaggio}

\affil[1]{Department of Physics and Center for Photonics and Nanoelectronics, Lehigh University, Bethlehem, PA 18015}

\affil[*]{Corresponding author: biaggio@lehigh.edu}

\begin{abstract}
We introduce  electrically-poled small molecule assemblies that can serve as the active electro-optic material in nano-scale guided-wave circuits such as those of the silicon photonics platform. These monolithic organic materials can be vacuum-deposited to homogeneously fill nanometer-size integrated-optics structures, and  electrically poled at higher temperatures to impart an orientational non-centrosymmetric order that remains stable at room temperature. An initial demonstration using the DDMEBT molecule and corona poling delivered a  material with the required high optical quality, an effective glass transition temperature of the order of $\sim 80$ $^\circ$C, and an electro-optic coefficient of $20$~pm/V.
\end{abstract}

\setboolean{displaycopyright}{true}

\begin{document}

\maketitle

Conjugated organic molecules, when combined into a material compatible with integrated optics, can contribute strong and fast nonlinear optical functionalities to state-of-the art integrated photonics platforms that already provide  unprecedented control of optical modes and their propagation. Silicon-organic-hybrid electro-optic modulators  have already been demonstrated  on the silicon photonics platform \cite{leuthold_silicon_2009,thomson_roadmap_2016,cheben_subwavelength_2018} using  electrically poled polymers \cite{palmer_high-speed_2014,lauermann_integrated_2016,heni_nonlinearities_2017,kieninger_ultra-high_2018}. These organic materials enable ultrafast electro-optic modulation because of their  low dielectric constant and  ultrafast electronic response of conjugated molecules.

However, even though electrically poled polymers have been developed over many decades and have record-high electro-optic coefficients \cite{Dalton03,Dalton10,Dalton11,kieninger_ultra-high_2018}, they are deposited from solution in a wet process that poses challenges towards a reproducible homogenous filling of nanostructures, especially when working  with nanoscale gaps of the order of $\sim 100$~nm or less  \cite{heni_nonlinearities_2017}. It is therefore important to investigate alternatives to polymers for the integration of organic materials and their active functionality into future photonics circuitry.

Here we propose vapor deposition of small molecules and subsequent electric poling of the resulting monolithic molecular assembly. 
Vacuum thermal evaporation of small organic molecules is currently widely used  in the the industrial production of organic light emitting displays \cite{Spindler17}. This dry,  solvent-free process  would enable the homogenous filling of  nanoscale integrated optical elements such as slot waveguides,  photonic bandgap structures, or subwavelength waveguides \cite{cheben_subwavelength_2018,Pan18,Govdeli18,Ghosh19}, and also offers a   nanometer-level control of thickness and precise mask-defined deposition areas.  

The electrically-poled small molecule assemblies (PSMA) discussed in this work are monolithic, dense organic materials consisting of small  dipolar molecules, and  fulfill the following requirements: (1) the molecules  sublimate without decomposition, enabling the use of vacuum thermal evaporation in manufacturing;   (2) the material obtained after deposition on any substrate is homogenous with a high optical quality; and  (3) the material can  be electrically poled at higher temperatures to impart a stable electro-optic activity. In these systems \cite{Biaggio22}, a higher molecular number density will at least partially compensate the necessarily lower optical nonlinearity of smaller molecules.

As a background, we note that some non-lattice-matched deposition of  small-molecule assemblies with orientational order  has been demonstrated earlier, but   did not generally involve molecules with large dipole moments or  deliver  electro-optic activity  \cite{Yokoyama09,Yokoyama10,zhu_surface_2011,dawson_anisotropic_2011,ediger_anisotropic_2019}, or else relied on directional hydrogen-bonding sites that required specially prepared  substrates \cite{Cai99}.  Vapor-transport deposition of crystalline films has also  been experimented with \cite{burrows_organic_1995,forrest_intense_1996,baldo_organic_1998}, but cannot be readily adapted to growing a single crystal over millimeter distances inside nano-scale structures.

\begin{figure}[h!]
\centerline{ \includegraphics[width= \linewidth]{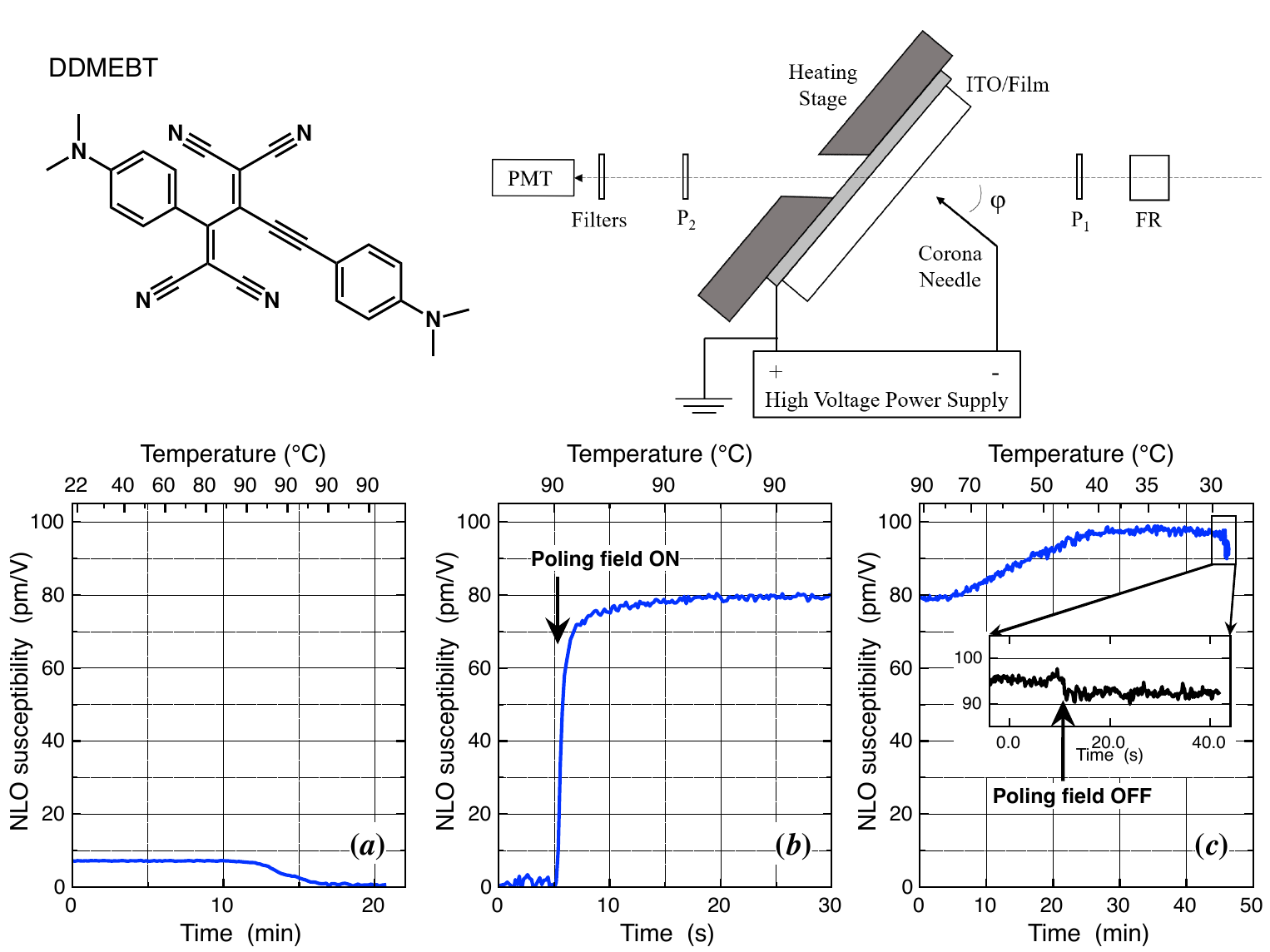}}
  \caption{ Top: The   DDMEBT molecule, with chemical formula  C$_{26}$H$_{20}$N$_6$, and full name  [2-[4-(Dimethylamino)phenyl]-3-([4(dimethylamino)phenyl]- ethynyl)-buta-1,3-diene-1,1,4,4-tetracarbonitrile)] \cite{esembeson_high_2008,michinobu_donor-substituted_2005},   and a scheme of the corona poling apparatus used in this work. Bottom: 
Poling procedure performed with a poling voltage of 5.5~kV on a  250~nm thick DDMEBT film. The plots give  the  second order nonlinear optical susceptibility of the film observed during the experiment. The bottom horizontal axis in every plot is time (note the different scale of the middle plot and the inset in the third plot). The tic marks on the top horizontal axes of every plot give the sample temperature at the corresponding moment in time.}
  \label{setupandpoling}
\end{figure}

For this demonstration, we used a material  (DDMEBT \cite{esembeson_high_2008}, see Fig.~\ref{setupandpoling}) that already fulfills requirements~(1) and~(2) in the above list: it deposits well using vacuum thermal evaporation \cite{esembeson_high_2008} and leads to a high optical-quality, homogenous material \cite{esembeson_high_2008,koos_all-optical_2009,scimeca_vapor_2009, Biaggio22}. In this work, we  show that it also fulfills requirement number (3). DDMEBT is a compact donor-acceptor substituted molecule with a high specific third-order polarizability \cite{erickson_optimum_2016},  and it has already been shown to homogeneously fill the gap of a slot waveguide used for  all-optical switching \cite{koos_all-optical_2009,scimeca_vapor_2009}.  This molecule  is also interesting for a PSMA because it  has a dipole moment of 11~Debye \cite{bures_solvatochromism_2011}, and a second order nonlinear optical polarizability  estimated to be of the order of $35\times10^{-39}$~m$^4/$V~ \cite{dallachiesa_vapor_2021}.

We deposited the DDMEBT material on glass substrates coated with Indium Tin Oxide (ITO) using vacuum molecular beam deposition at a pressure of   $10^{-5}$ Torr, with a deposition rate of  $\sim 1$~nm/min. Films with thicknesses ranging from 100~nm to 750~nm were then  poled by applying an electric field at higher temperatures using corona poling \cite{giacometti_corona_1992,eich_corona_1989}. The second order nonlinear optical susceptibility of the films, which directly relates to the degree of orientational ordering, was determineds using  second harmonic generation (SHG) before, during, and after corona poling.

For our SHG experiments we used 1~ps duration laser pulses at an off-resonant wavelength of 1500~nm and a repetition rate of 1~kHz, produced by a TOPAS (Traveling Wave Optical Parametric Amplifier System) from Light Conversion, pumped by a Clark MXR Ti:Sapphire amplifier (CPA-2001). The sample was mounted on a rotation stage, and its second-order nonlinear optical susceptibility was obtained by comparing, in a Maker fringes experiment  \cite{jerphagnon_maker_1970, herman_maker_1995},  the incidence-angle dependence of the SHG signal of the organic films to that of a 2 mm thick quartz crystal  (nonlinear optical susceptibility of $\chi^{(2)}_{111} = 0.6$ pm/V, corresponding to a ``d-coefficient'' $d_{111}=0.3$ pm/V \cite{Eckardt90}). 

SHG characterization at room temperature showed that as grown DDMEBT films already have a slight preferential molecular orientation that breaks symmetry in a direction perpendicular to the substrate and induces a small second-order nonlinear optical susceptibility \cite{erickson_-situ_2018}. This is a characteristic  of some vapor deposited organic glasses  \cite{lin_anisotropic_2004, zhu_surface_2011,dawson_anisotropic_2011, dawson_molecular_2012, lyubimov_orientational_2015, ediger_vapor-deposited_2014, ediger_anisotropic_2019}, where an orientational order  arises from the assembly process of the molecules at the vacuum-film interface \cite{zhu_surface_2011, lyubimov_orientational_2015}. Our thickness-dependent SHG measurements imply  that this  orientational order is  built into the DDMEBT film throughout its whole thickness. %We are not aware of any previous observation of this effect for electro-optic materials consisting of strongly dipolar molecules.

Electrical poling was done  in a corona poling set-up (see Fig.~\ref{setupandpoling}) with in-situ SHG characterization.   Fig.~\ref{setupandpoling} shows plots of the second-order susceptibility of the organic film, as it is continually observed during the whole procedure. Fig.~\ref{setupandpoling}($a$) is for the initial heating of the film  between room temperature and $90^\circ$C, using a linear  ramp of $5^\circ$C per minute. The  second order susceptibility already observed at toom temperature is the intrinsic one   imposed during vacuum deposition. It then starts slowly decreasing above $70^\circ$C as the intrinsic orientational order  is being thermally randomized, and vanishes completely after a few minutes at  $90^\circ$C (see Fig.~\ref{setupandpoling}$a$). This behavior is characteristic of an effective glass-transition temperature, an interpretation supported by the fact that the as-grown material is an amorphous molecular assembly with properties similar to previously investigated vapor-deposited organic glasses \cite{lin_anisotropic_2004, dawson_anisotropic_2011, dawson_molecular_2012, lyubimov_orientational_2015, ediger_vapor-deposited_2014, ediger_anisotropic_2019}. 

Once the intrinsic preferential orientational order completely disappeared at $90^\circ$C, a voltage of $5.5$ kV was applied to the corona needle. This  caused the second-order susceptibility (Fig.~\ref{setupandpoling}$b$) to  rise to a value of $\chi^{(2)} \sim 80$ pm/V within a fraction of a second.  The film is then  cooled back down while keeping the applied electric field constant (Fig.~\ref{setupandpoling}$c$). Interestingly, during the slow cooling process (about 40 minutes to reach 30$^\circ$C), the second-order susceptibility, and therefore the orientational order, continued to increase, to then reach  a final value of $\sim 100$ pm/V that remained stable at room temperature. 

Switching off the corona poling voltage at this point only caused a small drop in second-order susceptibility, consistent with the third-order nonlinear optical process of electric-field-induced SHG, as described by
\begin{equation}
	\frac{SHG(E=E_c)}{SHG(E=0)} =  \left[ 1 +  \frac{\chi^{(2)}_{111} }{\chi^{(3)}_{1111} E_c} \right]^2, \label{ElectricFieldDetermination}
\end{equation}
where $E_c$ is the effective electric field inside the material, the left-hand side is the ratio between the SHG powers measured with and without the corona poling voltage, $\chi^{(3)}_{1111} = (2 \pm 1) \times 10^{-19} {\rm \ m}^2{\rm V}^{-2}$ is the third-order susceptibility of DDMEBT at a wavelength of 1.5 $\mu$m \cite{esembeson_high_2008,beels_compact_2012}, and  $\chi^{(2)}_{111}$ is the second order susceptibility determined in this work. From this we can estimate the actual electric field that is present inside the film in our experiments to be of the order of  $E_c =  50 {\rm \ kV/mm } = 50 {\rm \ V/}\mu{\rm m}$ with an uncertainty of $\pm 50$\%.

The continued increase of the preferential orientational order when the films are cooled under the applied electric field is  not  generally  observed  when similar procedures are applied to molecules in a solid solution such as a polymer-based guest-host systems. However, our material is made up of a dense monolithic assembly of small molecules. Therefore, the molecular dipoles cannot simply reorient, there also needs to be a rearrangement of the molecular packing, which in this case appears to continue to  occur, slowly, at lower temperatures. While more experiments will be needed to completely characterize this process, we can tentatively speculate that the DDMEBT molecular structure may play a role. This molecule consists of two parts that are almost perpendicular to each other. The corresponding rotation angle around the single bond is of 96.7 degrees \cite{Michinobu06}, close to perpendicular. This may give this molecule some flexibility in adapting to its environment upon thermal perturbations. This very flexibility likely aids towards forming an amorphous solid state upon deposition, and it is reasonable to assume that it also plays  a role  in the continued growth of the orientational order of the DDMEBT solid state state material that is observed (see Fig.~\ref{setupandpoling}$c$) at temperatures slightly below the poling temperature.

We repeated the experiment several times and confirmed that the poling procedure described above also works at slightly lower or larger poling temperatures, with longer times for the randomization of the intrinsic order at the lower temperatures,  a sub-second build-up of the electric-field-induced ordering when applying the corona voltage, and an additional slower improvement of the order later on. Poling results in different samples are  given in Table~\ref{poled_films}. The sample-to-sample variaions in the electro-optic coefficient shown in this table are generally small, and they could also be attributed to slight changes in the corona poling setup (such as the type of needle or the gap between the tip of the needle and the film). The one outlier with the significantly lower value belongs to an earlier experiment.

\begin{figure}[t]
\centerline{ \includegraphics[width= \linewidth]{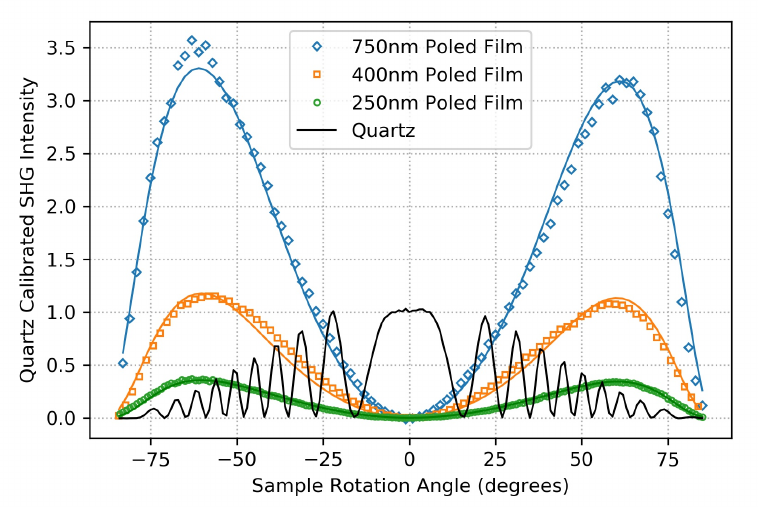}}
  \caption{Incidence angle dependence of SHG  from three different PSMA films and the quartz reference. Both fundamental and second-harmonic are polarized parallel to the incidence plane.  The solid curves are theoretical fits  with a second-order susceptibility for the films of $\chi_{111}^{(2)}~=~100$ pm/V, where the 1-axis is perpendicular to the substrate} 
  \label{fig:filmsandquartz}
\end{figure}

We have not seen any detrimental effects of the poling procedure on the   vapor-deposited material, which remains of  high optical quality,   with no signs of crystallization (as confirmed by light scattering and x-ray diffraction). In addition to this, SHG characterization was  repeated at different times after deposition (up to a year), and confirmed  that the preferential molecular orientation remains stable. The results of a full Maker fringes  SHG measurement of three poled films  with different thicknesses and of the quartz reference is shown in Fig.~\ref{fig:filmsandquartz}.  The  second-order nonlinear optical susceptibility  of the poled films was obtained by comparing the SHG power from  the films and from the quartz reference while taking into account all projections of the optical fields, Fresnel transmission coefficients, and the DDMEBT refractive indices of 1.8 at 1550 nm and 2.0 at 750 nm (see Ref.~\cite{esembeson_high_2008} for the refractive index dispersion in DDMEBT). The result obtained from the data in Fig.~\ref{fig:filmsandquartz} is   $\chi^{(2)}_{111} = 100\pm30$~pm/V for all three films with different thicknesses, showing that the orientational order is maintained throughout the bulk of the material and is reproducible between different samples.

In organic materials, one expects a good correspondence between the susceptibilities for SHG and for electro-optics, leading to the following relationship between the second-order susceptibility for SHG and the standard electro-optic coefficient $r_{ijk}$,
\begin{equation}
\chi_{ijk}^{(2)}(-\omega,\omega,0) = - \frac{1}{2} n_i^2 n_j^2 r_{ijk} , \label{eq:rChi}
\end{equation} 
where $n_1=n_2=n_3 = 1.8$ is the refractive index of the  DDMEBT material at 1.5 $\mu$m \cite{esembeson_high_2008}. Here we can use the isotropic refractive index determined for the as-grown material because the preferential (incomplete) alignment that we obtain after poling has a negligible influence on the linear optical properties.
Applying  this expression to  the experimental value of {$\chi^{(2)}=100 \pm 30$~pm/V} we find that the corresponding electro-optic coefficient for a pure electronic response and ultra-fast electro-optic modulation is $r_{111} = 20$~pm/V.   As a reference, this value is  only $\sim 30$\% lower than the electro-optic coefficients of  standard electro-optic crystals such as ferroelectric KTP (KTiOPO$_4$) or LiNbO$_3$.

\begin{table}[h]
\centering
\caption{Electro-optic coefficients  obtained in  DDMEBT films after applying the high-temperature corona poling procedure shown in Fig.~\ref{setupandpoling} (from SHG characterization and Eq.\ref{eq:rChi}).}
\begin{tabular}{ccc}
\hline
Film   & Corona    & $r_{111}$  \\ 
Thickness (nm) & Voltage (kV)  & (pm/V) \\ 
\hline
250     & -5.5                 & 20  \\
300     & -5.0                  & 13  \\
350     & -5.0                   & 18  \\
420     & -5.5                   & 22  \\
750     & -5.5                 & 18  \\
\hline
\end{tabular}
\label{poled_films}
\end{table}

The size of the electro-optic coefficient of the PSMA demonstrated here could already enable a ultra-high-speed electro-optic modulator on the silicon photonics platform using a millimeter-long slot-waveguide \cite{alloatti_100_2014}. Still, we should point out that in our experiments the poling electric field  was limited to an estimated $ 50 {\rm \ kV/mm } = 50 {\rm \ V}/\mu{\rm m}$, but the use of up to five times higher poling electric fields is possible  \cite{palmer_high-speed_2014, lauermann_integrated_2016, kieninger_ultra-high_2018}. In addition,  this PSMA value of $r_{111} = 20$~pm/V  was obtained 
using an off-the-shelf molecule that was not explicitly optimized for electro-optics or for this fabrication method. There is therefore  still a strong potential for future improvements. 

In conclusion, the present demonstration of an electrically poled monolithic small-molecule assembly shows that the necessarily weaker second-order polarizability of a small molecule that can sublimate without decomposition can be compensated by the larger molecular number-density in a monolithic PSMA. An electro-optic coefficient of 20 pm/V was demonstrated, and there is potential for significant  improvements after  research on the optimization of molecular structure and properties. 

The ability to integrate the organic material with existing state-of-the-art guided-wave  technology using vacuum thermal evaporation instead of a  wet process of deposition from solution enables a better control of the deposition process and homogeneous incorporation of the organic material using a technique that can be easily scaled from the laboratory  to industrial production  \cite{Spindler17}.  

PSMAs can therefore become a valuable alternative paradigm towards  adding a second-order nonlinear optical functionality to integrated optics platforms, not only for electro-optic effects, but also for such applications as optical parametric generators or single photon or photon-pair sources.

\begin{backmatter}

\bmsection{Disclosures} The authors declare no conflicts of interest.

\end{backmatter}

% Bibliography
\bibliography{DDMEBT}

% Full bibliography added automatically for Optics Letters submissions; the following line will simply be ignored if submitting to other journals.
% Note that this extra page will not count against page length
\bibliographyfullrefs{DDMEBT}

%Manual citation list
%\begin{thebibliography}{1}
%\bibitem{Zhang:14}
%Y.~Zhang, S.~Qiao, L.~Sun, Q.~W. Shi, W.~Huang, %L.~Li, and Z.~Yang,
 % \enquote{Photoinduced active terahertz metamaterials with nanostructured
  %vanadium dioxide film deposited by sol-gel method,} Opt. Express \textbf{22},
  %11070--11078 (2014).
%\end{thebibliography}

% Please include bios and photos of all authors for aop articles
\ifthenelse{\equal{\journalref}{aop}}{%
\section*{Author Biographies}
\begingroup
\setlength\intextsep{0pt}
\begin{minipage}[t][6.3cm][t]{1.0\textwidth} % Adjust height [6.3cm] as required for separation of bio photos.
  \begin{wrapfigure}{L}{0.25\textwidth}
    \includegraphics[width=0.25\textwidth]{john_smith.eps}
  \end{wrapfigure}
  \noindent
  {\bfseries John Smith} received his BSc (Mathematics) in 2000 from The University of Maryland. His research interests include lasers and optics.
\end{minipage}
\begin{minipage}{1.0\textwidth}
  \begin{wrapfigure}{L}{0.25\textwidth}
    \includegraphics[width=0.25\textwidth]{alice_smith.eps}
  \end{wrapfigure}
  \noindent
  {\bfseries Alice Smith} also received her BSc (Mathematics) in 2000 from The University of Maryland. Her research interests also include lasers and optics.
\end{minipage}
\endgroup
}{}

\end{document}